# Category decoding of visual stimuli from human brain activity using a bidirectional recurrent neural network to simulate bidirectional information flows in human visual cortices

**Kai Qiao[1], Jian Chen[1], Linyuan Wang[1], Chi Zhang[1], Lei Zeng[1], Li Tong[1], Bin Yan[*1]**

[1] National Digital Switching System Engineering and Technological Research Center, Zhengzhou, China

**\* Correspondence:**
Dr. Bin Yan
ybspace@hotmail.com

**Abstract**

Recently, visual encoding and decoding based on functional magnetic resonance imaging (fMRI) have realized many achievements with the rapid development of deep network computation. Despite the hierarchically similar representations of deep network and human vision, visual information flows from primary visual cortices to high visual cortices and vice versa based on the bottom-up and top-down manners, respectively. Inspired by the bidirectional information flows, we proposed a bidirectional recurrent neural network (BRNN)-based method to decode the categories from fMRI data. The forward and backward directions in the BRNN module characterized the bottom-up and top-down manners, respectively. The proposed method regarded the selected voxels of each visual cortex region (V1, V2, V3, V4, and LO) as one node in the sequence fed into the BRNN module and combined the output of the BRNN module to decode the categories with the subsequent fully connected layer. This new method allows the efficient utilization of hierarchical information representations and bidirectional information flows in human visual cortices. Experiment results demonstrated that our method improved the accuracy of three-level category decoding than other methods, which implicitly validated the hierarchical and bidirectional human visual representations. Comparative analysis revealed that the category representations of human visual cortices were hierarchical, distributed, complementary, and correlative.

## 1    Introduction

In neuroscience, the mechanisms by which visual stimuli are encoded by neurons and visual stimuli information is decoded from neuronal activities have yet to be elucidated. Functional magnetic resonance imaging (fMRI) reflects brain activities effectively; hence, visual decoding computation models based on fMRI data have attracted considerable attention (Haynes and Rees, 2006; Norman et al., 2006; Naselaris et al., 2011; Nishimoto et al., 2011). In contrast to visual encoding computation models (Kay et al., 2008; Styves and Naselaris, 2017; Wen et al., 2018) that predict brain activities in response to visual stimuli, inverse decoding computation models (Kamitani and Tong, 2005; Horikawa et al., 2013; Li et al., 2018; Papadimitriou et al., 2018) predict information about visual stimuli through brain activities. Classification, identification, and reconstruction of image stimuli based on fMRI data are three main means of visual decoding. Compared with identification and reconstruction, category decoding is common and feasible in brain decoding, considering that the identification is limited to fixed dataset and the fine reconstruction is limited to simple image stimuli.

The category decoding of visual stimuli can be summarized into three main methods: 1) classifier-based methods, 2) voxel pattern template matching-based methods, and 3) feature pattern template matching-based methods. Classifier-based methods accomplish category decoding by training a statistical linear or nonlinear classifier to directly learn the relationship between specific voxel patterns and the category to be decoded. Previous works (Cox and Savoy, 2003) employed several linear support vector machine (SVM) classifiers (Chang and Lin, 2011) to discriminate the distributed voxel patterns evoked by each category. In addition to SVM, various classifiers, including fisher classifier and k-nearest neighbors, have been used (Misaki et al., 2010; Song et al., 2011) in category decoding. Wen et al. (2017) employed the pre-trained last classifier of the convolutional neural network (CNN) (LeCun et al., 1998) in category decoding. Voxel pattern template matching-based methods accomplish category decoding by computing the correlation between voxels to be decoded and the voxel pattern template of each category, and the maximum correlation indicates the category. This kind of method (Sorger et al., 2012) aims to construct a voxel pattern template for each category. Haxby et al. (2001) directly constructed voxel pattern templates by averaging voxels of those samples belonging to the same category. Meanwhile, Kay et al. (2008) constructed an encoding model and then obtained the voxel pattern template of each category by averaging the predicted voxel pattern based on the encoding model and those samples belonging to the corresponding category. Feature pattern template matching-based methods accomplish category decoding by training the mapping from voxels to specific feature representations, comparing the predicted features with a feature pattern template of each category, and finally selecting the category that demonstrates the maximum correlation. The third manner depends on the feature bridge; hence, the mapping from voxels to feature representations plays an important role. Horikawa and Kamitani (2017) and Wen et al. (2018) constructed a feature pattern template for each category by averaging the predicted CNN features of all image stimuli belonging to the same category. Among these studies, those on hierarchical CNN features have received much attention (Agrawal et al., 2014; Güçlü and van Gerven, 2015).

In the human vision system, bidirectional information flows in visual cortices during the processing of visual information. On the one hand, visual information flows from primary visual cortices to high-level visual cortices to extract high-level semantic content through hierarchical representations. In this way, people can understand visual stimuli based on the so-called "bottom-up visual mechanism" (Logothetis and Sheinberg, 1996). On the other hand, visual information can also flow from high-level visual cortices to primary visual cortices on the basis of the visual attention mechanism (Ungerleider and G, 2000; Carrasco, 2011). In this way, people can pay attention to regions of interest on the basis of the "top-down visual mechanism" (Beck and Kastner, 2009; McMains and Kastner, 2011). Obviously, visual vortices are dependent, and the bottom-up and top-down manners allow the bidirectional information flows to carry semantic knowledge. Therefore, maximizing the each visual cortex and the bidirectional information flows in visual cortices is of great significant for the category decoding.

The current three types of methods employed specific visual cortices or parallelly processed all visual cortices, which can maximize each visual cortex, but not the bidirectional information flows in visual cortices. Therefore, we introduced the bidirectional information flows into the decoding model to improve the decoding performance. Unlike feedforward neural networks, recurrent neural networks (RNNs) (Mikolov et al., 2010; Graves et al., 2013; LeCun et al., 2015) can perform extremely well on temporal data and have realized great achievements in sequence modeling. The general RNN architecture is assumed to only have one directional connection from past to future (or from left to right) input nodes. Bidirectional recurrent neural networks (BRNNs) (Schuster and Paliwal, 1997; Schmidhuber, 2015) split the neurons of a regular RNN into two directions: one for positive time direction and another for negative time direction. With the two directions, input information from the past and future of the current time frame can be used. Inspired by BRNN, we can regard the bidirectional information flows between visual cortices as the temporal sequence. Therefore, we propose to feed each visual cortex as one node of the sequence into the bidirectional long short-term memory (LSTM) (Hochreiter and Schmidhuber, 1997; Sutskever et al., 2014) module, a prevailing type of RNN architecture. In this way, the output of the bidirectional LSTM module can be regarded as the representations of the bottom-up and top-down manners. The category can be predicted with subsequent fully connected layers by combining the bidirectional representations.

In this study, our main contributions are as follows: 1) we analyzed the drawbacks of current decoding methods from the perspective of bidirectional information flows (bottom-up and top-down visual mechanisms); 2) we proposed to employ the BRNN module to simulate the bidirectional information flows for the category decoding of visual stimuli; and 3) we analyzed and concluded that the category representations of human visual cortices were hierarchical, distributed, complementary, and correlative.

## 2 Materials and Methods

### 2.1 Experimental data

We employed the dataset from previous studies (Kay et al., 2008; Naselaris et al., 2009). The 4T INOVA MRI system was used to acquire blood-oxygenation-level dependent (BOLD) sensitive functional images for two human subjects (S1 and S2). The single-shot gradient EPI sequence with a repetition time (TR) of 1 second and isotropic voxel size of 2×2×2.5 mm$^3$ was employed. Five regions of interest (V1, V2, V3, V4, and LO) from low-level to high-level visual cortices were included in the dataset. The dataset was divided into two sets: training and validation. The training and validation sets included 1750 and 120 grayscale natural images randomly selected from a database, respectively. The acquired data was subjected to pre-processing, which included motion correction, registration with the anatomical scan, and detrended and high-pass-filtered dealing with the hemodynamic response function issue within statistical parameter mapping software. In addition, the detailed information about the fMRI data is provided in previous studies (Kay et al., 2008; Naselaris et al., 2009), and the public dataset can be downloaded from *http://crcns.org/data-sets/vc/vim-1*. In addition to the image stimuli and corresponding fMRI data, five persons hired in this study manually labeled the 1870 images according to there-level categories and final labels were obtained through voting. As shown in the Figure 1, the dataset with three-level categories (5, 10, and 23 categories) can comprehensively validate the decoding method from coarse to fine grains.

### 2.2 Overview of the proposed method

To introduce the bidirectional information flows into the decoding method to improve the decoding performance, we employed a BRNN-based method to simulate the bottom-up and top-down manners in the human vision system. In this way, not only information in each cortex but also bidirectional information flows in visual cortices can be used in the decoding method. As shown in the Figure 2, the proposed model included the encoding and decoding parts. For the encoding part, we can obtain corresponding features given image stimuli based on the prevailing pre-trained ResNet-50 (He et al., 2016) model and employed these features to fit each voxel to construct the voxel-wise encoding model. According to the fitting performance, we can measure the importance of each voxel in all visual cortices. We selected the fixed small number of voxels for each visual cortex (V1, V2, V3, V4, and LO) and fed them into the subsequent decoding part. In this way, those voxels with higher predictive correlation can be selected to prevent overfitting when training the decoding mapping. For the decoding part, we constructed a bidirectional LSTM module and employed the selected voxels of each visual cortex as the five nodes of sequence input to utilize both hierarchical visual representations and bidirectional information flows in visual cortices. Finally, we combined the output of bidirectional LSTM module features as the input of last fully connected layer to predict the category.

Section 2.3 introduces how to construct a visual encoding model based on hierarchical feature representations through pre-trained CNN architecture. Section 2.4 demonstrates how to use a BRNN to simulate the bidirectional information flows in visual cortices to decode the category.

### 2.3 Visual encoding with hierarchical feature representations

**Extracting hierarchical visual features based on pre-trained ResNet-50.** Many works (Agrawal et al., 2014; Yamins et al., 2014; Güçlü and van Gerven, 2015) indicated that the hierarchical features of the pre-trained CNN model performed similar representations with the human vision system, and the visual encoding based on those hierarchical features improved the encoding performance. In this study, we used the prevailing deep network ResNet-50 to extract hierarchical features for successive visual encoding. The pre-trained ResNet-50 can recognize 1000 types of natural images (Russakovsky et al., 2015) with state-of-the-art performance, which demonstrated that the network possessed rich and powerful feature representations.

In ResNet-50, 50 hidden layers were stacked into a bottom-up hierarchy. Aside from the first convolutional layer, four modules (16 residual blocks with each block mainly comprising three convolutional layers) and the last fully connected layers were included. Compared with AlexNet (Krizhevsky et al., 2012) in previous studies, ResNet-50 was much deeper and contained more fine-grained hierarchical representations. As shown in the Table 1, we selected features including outputs of the last avgpooling operation and 16 residual blocks for visual encoding. In this way, we can extract seventeen types of features for each image stimulus and prepare 1870 features-voxels samples. In the experiment, we used the pre-trained model from the prevailing PyTorch toolbox (Ketkar, 2017).

**Voxel-wise encoding voxels based on sparse regression.** For each layer, a linear regression model was defined to map the feature vector $X$ to each voxel $y$, and it is expressed as follows:

$$\mathbf{y} = \mathbf{X}\mathbf{w}, \quad (1)$$

where $y$ is an $m$-by-1 matrix and $X$ is an $m$-by-$n$ matrix, where $m$ is the number of training images and $n$ is the dimensionality of the feature vector. $w$, an $n$-by-1 matrix, is the unknown weighting vector to solve. Table 1 presents the dimensionality for each selected feature vector. The number of training samples $m$ is considerably less than the number of features $n$ in the feature vector, which complicates training the mapping from the high dimensionality of the feature vector to voxels with relatively few training samples. Thus, we assumed that each voxel can be characterized by a small number of features in the feature vector and regularized $w$ sparse to prevent overfitting when training the mapping. On the basis of the above assumption, the major problem can be expressed as follows:

$$\min_{w} \|w\|_0 \quad \text{subject to} \quad \mathbf{X}\mathbf{w} = \mathbf{y}. \quad (2)$$

In this study, we employed the regularized orthogonal matching pursuit (ROMP) (Needell and Vershynin, 2010), a sparse optimization method, to fit the voxel pattern. ROMP adds orthogonal item and group regularization based on the matching pursuit algorithm (Mallat and Zhang, 1993). Algorithm steps are detailed in the study by Needell and Vershynin (2010). We performed voxel-wise encoding using features of each layers and selected the optimal performance as the final features for the encoding model. Figure 3 presents the encoding performance of the selected voxels for each visual cortex. In the figure, the features of the optimal layer are marked in the "circle" according to the encoding performance. Finally, we selected the top 100 voxels in each visual cortex (V1, V2, V3, V4, and LO) according to the predicting performance, and a total of 500 voxels were selected for the next category decoding. On the basis of the encoding model, the dimensionality of voxels for each cortex was reduced to the small and fixed number, whereas valuable information of voxels was reserved. In addition, the low-level features were fit for encoding the primary visual cortex, and high-level features were fit for encoding high-level visual cortex, which agreed with a previous study (Wen et al., 2018).

### 2.4 Modeling of bidirectional information flows in visual cortices with the BRNN

**LSTM module.** Based on the traditional RNN module, LSTM and gated recurrent unit (GRU) (Cho et al., 2014) have been widely used as an RNN module. In this study, we employed the LSTM and found no benefit with GRU. LSTM is normally augmented by recurrent gates called "forget" gates and can prevent backpropagated errors from vanishing or exploding. LSTM can learn tasks that require memories of events that previously occurred. As shown in the Figure 4, LSTM added one state variable $c_t$ compared with the traditional RNN module, and $c_t$ can render the network include long-term or short-term memory as needed. LSTM includes three gates ("forget," "input," and "output" gates), which depend on previous state $h_{t-1}$ and current input $x_t$. The "forget" gate can control how much to forget previous information according to $f_t$ computed through Equation (3), where $\sigma$ represents the sigmoid function to restrict $f_t$ from 0 to 1. The "input" gate can control how much to feed current input $x_t$ into the computation according to $i_t$ computed through Equation (4). The "output" gate can control to output how much information according to $o_t$ computed through Equation (5).

$$f_i = \sigma(W_f \cdot [h_{t-1}, x_t] + b_f), \quad (3)$$

$$i_i = \sigma(W_i \cdot [h_{t-1}, x_t] + b_i), \quad (4)$$

$$o_i = \sigma(W_o \cdot [h_{t-1}, x_t] + b_o). \quad (5)$$

On the basis of the three gates, LSTM can compute the state $c_t$ and $h_t$, which is also the output of the current computation.

$$\tilde{c}_t = tanh(W_c \cdot [h_{t-1}, x_t] + b_c), \quad (6)$$

$$c_t = f_t \cdot c_{t-1} + i_t \cdot \tilde{c}_t, \quad (7)$$

$$h_t = o_t \cdot tanh(c_t). \quad (8)$$

**Category decoding architecture.** The connection in the RNN module usually only has one direction (from left to right), and the BRNN adds the other direction (from right to left) to render the module bidirectional. Basing from the LSTM module and selected voxels, we presented the category decoding architecture based on the bidirectional LSTM module. As shown in the Figure 5, the sequence length is five and the 100 selected voxels from each visual cortex are regarded as one node of the input sequence fed into the bidirectional LSTM module. The directions from left to right and right to left characterize the bottom-up and top-down manners in the human vision system, respectively. Then, the proposed method combined the output of each node and fed it into the successive fully connected layer to predict the category. Figure 5 also shows that each output features of one node are related to left low-level visual cortices and right high-level visual cortices. In this way, the relationship between cortices is used in the decoding model. The proposed method can be trained in an end-to-end manner with similar algorithms as standard RNN. Through training, the bidirectional information flows, including category information, can be mined on the basis of training samples. The dataset included three-level labels, and the number of nodes in the last fully connected layer can be changed to fit the 5-, 10-, and 23-category decoding.

## 3 Results

### 3.1 Conventional classifiers

We selected some widely used classifiers, including decision tree (DR), random forest (RF), AdaBoost (AB) and SVM. We performed the three-level category (5, 10, and 23) decoding on the basis of these conventional classifiers. To take an example from Figure 6 for the 23-category decoding, we can observe that these conventional methods are all above the chance with a single visual cortex, and even primary visual cortex benefits the semantic category decoding. We also plotted the linear trend of decoding performance from low-level to high-level visual cortices and observed that the decoding improved. This result indicates that higher-level visual cortices obtain more semantic information. In addition, these methods can obtain better decoding performance when all visual cortices instead of a single visual cortex are used, and different visual cortices contain some specific representations. The results of other two levels (5 and 10) of category decoding can be found in Figures 7 and 8, which demonstrated a similar phenomenon.

### 3.2 Neural network and our proposed method

To validate our proposed method, as shown in the Figure 9, we compared our proposed method with the above conventional methods and fully connected NN method. The NN method employed similar network architecture except RNN module. The NN method performed better than the conventional methods because of the powerful nonlinear ability of the former. Our proposed method performed the best regardless of which level category decoding because it can utilize the data, including the bidirectional information flows. The performance of the proposed method reached about 60% for 5-category decoding, 50% for 10-category decoding, and 40% for 23-category decoding. Table 2 demonstrates the explicit quantity, and the proposed method obtained 5.00%, 6.66%, and 7.50% improvement than other methods for 5-, 10-, and 23-category decoding, respectively. Similar results for subject S2 can be found in Table 3.

## 4 Discussion

### 4.1 Hierarchical representations about the category in visual cortices

As shown in the Figures 6, 7, and 8, the linear trend of the decoding performance of many methods indicated that the decoding performance improved from low-level visual cortices to high-level visual cortices. As shown in the Figure 3, the low-level and high-level features are suitable to encode low-level and high-level visual cortices, respectively. Low-level and high-level features of deep network focused on detailed and abstract information, respectively. The above appearance indicates that the human vision system processes and extracts hierarchical representations through the bottom-up information flow and that higher-level visual cortices contain more semantic information about the category. Therefore, the category representations of human visual cortices are hierarchical.

### 4.2 Distributed representations about the category in visual cortices

According to the results regardless of how many categories to be decoded (Figures 6, 7, and 8) when using only one specific visual cortex for different classifiers, the category decoding performance of each visual cortex is above the chance, and even the primary visual cortex can contribute to category decoding, which can indicate that low-level visual cortices can contain abstract information about the category. Thus, we conclude that some semantic feedback can flow from high-level visual cortices into low-level visual cortices through the top-down information flows, except hierarchical representations based on the bottom-up information flows. Therefore, the category representations of human visual cortices are distributed.

### 4.3 Complementary representations about the category in visual cortices

Distributed representations about the category in visual cortices make different visual cortices contain some specific representations about the category. Comparison of a single visual cortex and all visual cortices revealed that the visual representations of all visual cortices improved the decoding performance. The improvement indicates that low-level visual cortices are responsible for low-level representations (edge, texture, and colour) and that high-level visual cortices are responsible for high-level representations (shape and object). These representations include different levels of information, all of which contribute to category decoding. Therefore, the category representations of human visual cortices are complementary.

### 4.4 Correlative representations about the category in visual cortices

Complementary representations about the category in visual cortices show that combining different levels of cortices benefits the category decoding. In addition, comparative analysis (Figure 9) on using the bidirectional information flows revealed about 5% improvement in category decoding when introducing the bidirectional information flows in visual cortices. Thus, we can conclude that the bidirectional information flows in human visual cortices carry some category information. Our proposed method employs a BRNN-based method to successfully characterize the bottom-up and top-down manner information flows. Except hierarchical, distributed and complementary representations about the category in visual cortices, each visual cortex is related to other visual cortices and not independent. Therefore, the category representations of human visual cortices are correlative.

### 4.5 Bottom-up and top-down visual mechanisms

The above analysis suggests that the category representations of human visual cortices are hierarchical, distributed, complementary, and correlative, which can be attributed to the bottom-up and top-down visual mechanisms in the human vision system. Hence, we assume that the bottom-up and top-down manners should receive more attention in the visual encoding and decoding. Current methods, such as prevailing CNN-based methods, are usually restricted to hierarchical representations. However, the CNN architecture is hard to simulate the top-down visual mechanism. Therefore, incorporating more human visual mechanisms to characterize human visual processing can benefit visual encoding while decode more accurate information about the presented stimuli.

### 5 Conclusion

In this study, we analyzed the drawbacks of current decoding methods from the perspective of the bidirectional information flows (bottom-up and top-down visual mechanisms). We regarded the selected voxels of each visual cortex region (V1, V2, V3, V4, and LO) as one node in the sequence and proposed to employ the BRNN to characterize bidirectional information flows in visual cortices for category decoding. We tested our proposed method on the dataset with three levels (5, 10, and 23) category labels. Experimental results demonstrated that our proposed method is capable of more accurate category decoding than other linear and nonlinear classifiers, thereby validating the significance of bidirectional information flows for category decoding. In addition, basing from experimental results, we analyzed and concluded the hierarchical, distributed, complementary, and correlative representations about categories in visual cortices.

## 6  Author Contributions

KQ contributed to all stages of the research project and writing; JC designed the procedures of overall experiment; LW contributed to the idea of decoding based on the BRN; CZ contributed to the implementation of the idea; LZ contributed to the preparation of the article, figures, and charts; LT introduced the perception of hierarchical, distributed, complementary, and correlative representations in visual cortices; and BY proposed the idea and writing.

## 7  Conflict of Interest

The authors declare that the research was conducted in the absence of any commercial or financial relationships that could be construed as a potential conflict of interest.

## 8  Funding

This work was supported by the National Key Research and Development Plan of China (No. 2017YFB1002502), the National Natural Science Foundation of China (No. 61701089), and the Natural Science Foundation of Henan Province of China (No. 162300410333).

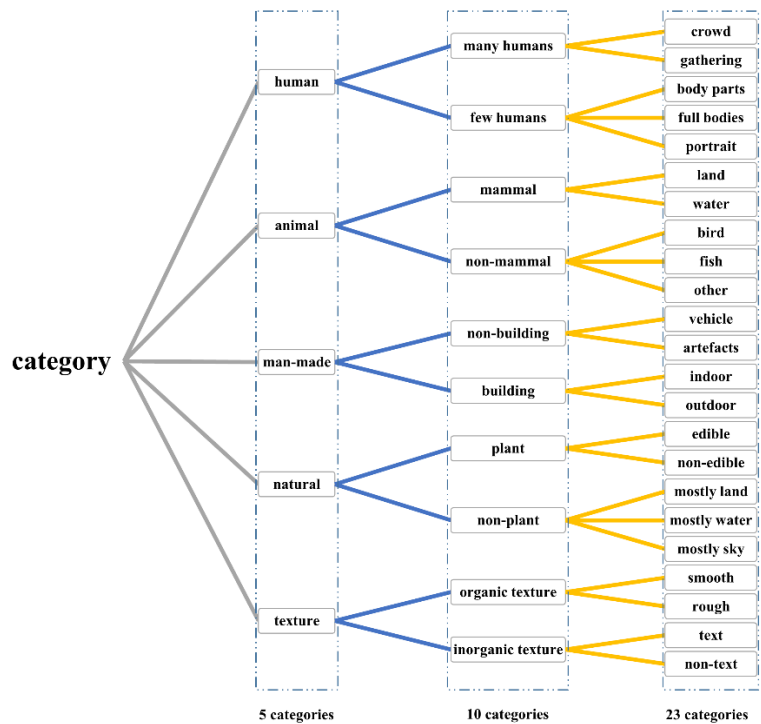

Figure 1. Three-level labels that have 5, 10, and 23 categories. Three-level categories were designed to validate methods according to different grains, which can make the comparison more persuasive.

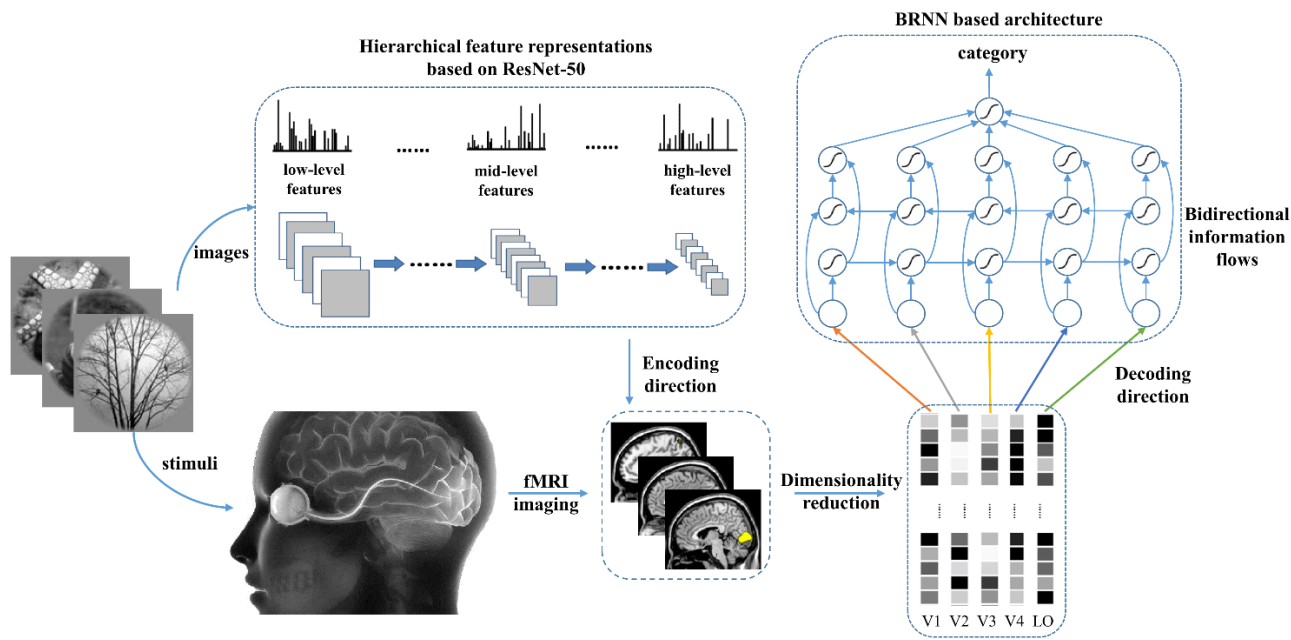

Figure 2. Proposed method. Hierarchical features in the deep network were used to predict voxel patterns in each visual cortex for the encoding direction. Based on the performance, the valuable voxels can be selected to reduce the dimensionality of voxels to a fixed number for each visual cortex. To predict the category, the voxel sequence comprising five voxels from five visual cortices is fed into the BRNN-based methods to extract semantic information from each visual cortex and the bidirectional information flows in visual cortices.

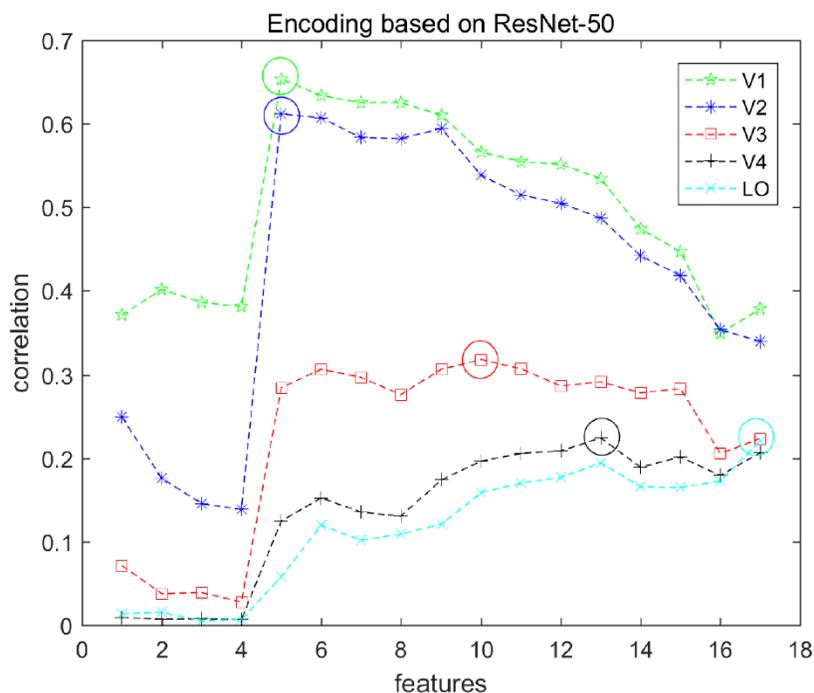

Figure 3. Encoding performance of each visual cortex based on ResNet-50 features. Seventeen types of features were used to encode each voxel in each visual cortex (V1, V2, V3, V4, and LO), and each node represents the average encoding performance of the top 100 voxels with higher correlation. Each color represents one type of visual cortex, and the corresponding "circle" indicates the optimal performance. In this way, the optimal features can be selected and the top 100 voxels were selected using optimal features for each cortex.

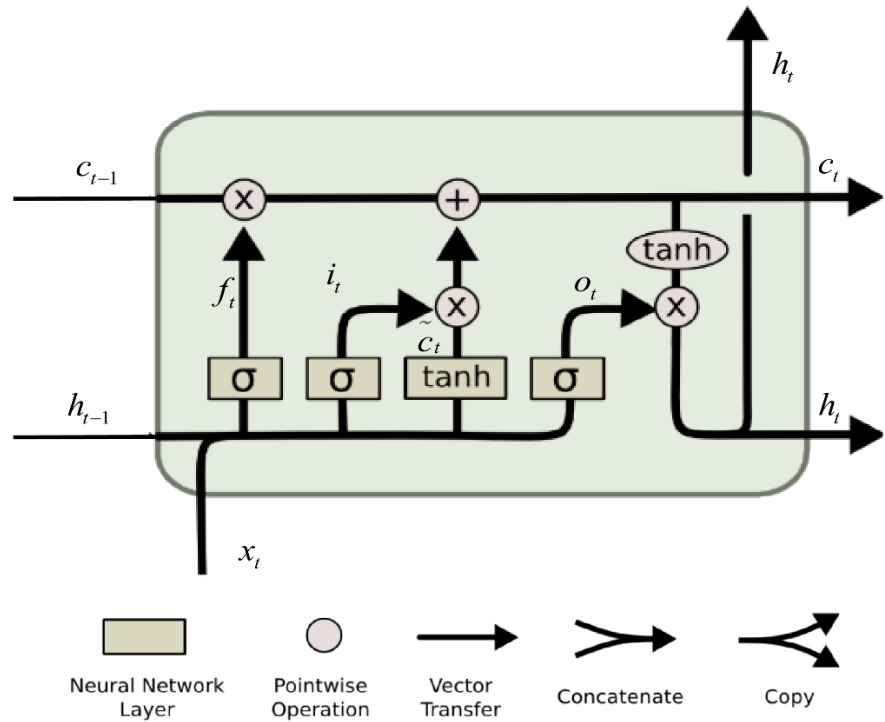

Figure 4. LSTM module for sequence modeling. Through the "forget" gate $f_t$, "input" gate $i_t$, and "output" gate $o_t$, the LSTM module can obtain the long-term and short-term memory. $c_t$ can be regarded as the input line and play an important role in the module. Current input $x_t$ feeds into the input line controlled by the previous state $h_{t-1}$ and itself, and can be used for the next state. In this way, LSTM can characterize each input node and the relationship between nodes in the sequence.

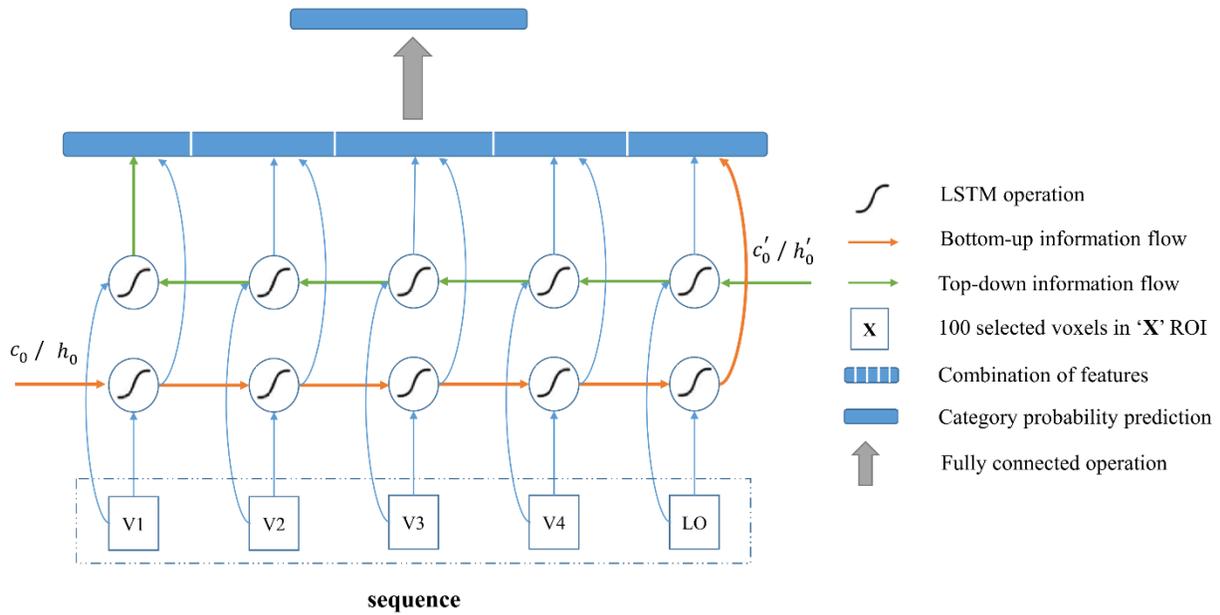

Figure 5. Category decoding model based on the BRNN module. All visual cortices were regarded as one sequence, and the BRNN module is especially good at sequence modeling. The red line indicates the bottom-up information flows, and the green line indicates the top-down information flows in visual cortices. Five channels represented the contribution of each visual cortex for the category decoding, and each channel contained the own information but also the bidirectional information from left low-level and right high-level visual cortices. The combination of features from five channels was used to predict category. In this way, information from each visual cortex and bidirectional information flows in visual cortices can both be used for the decoding.

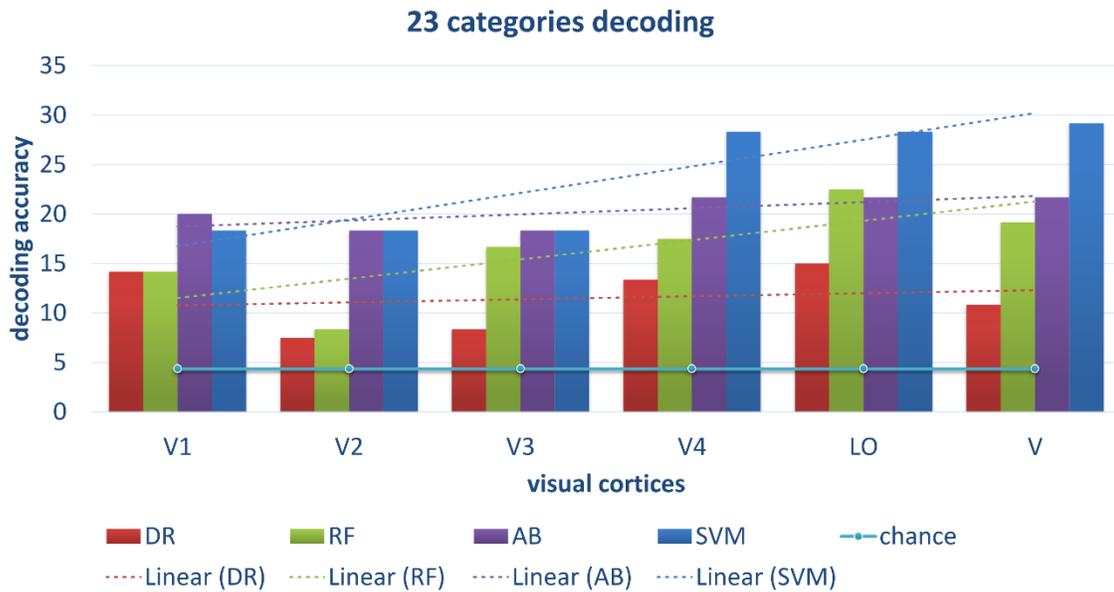

Figure 6. Decoding of 23 categories using conventional classifiers. Performances based on each visual cortex and all visual cortices ("V") when using different conventional classifiers were presented. The distributed, hierarchical, and complementary representations for the semantic category in the human vision system can be observed (detailed analysis in the Discussion part).

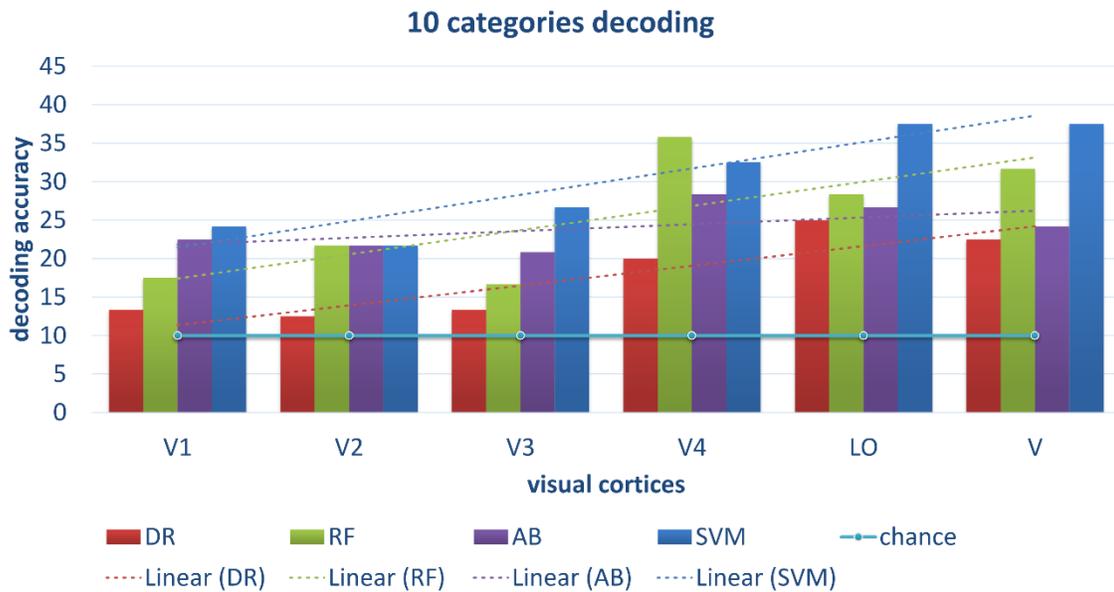

Figure 7. Decoding of 10 categories using conventional classifiers. Results demonstrated the phenomenon: distributed, hierarchical, and complementary representations about the semantic category in human visual cortices.

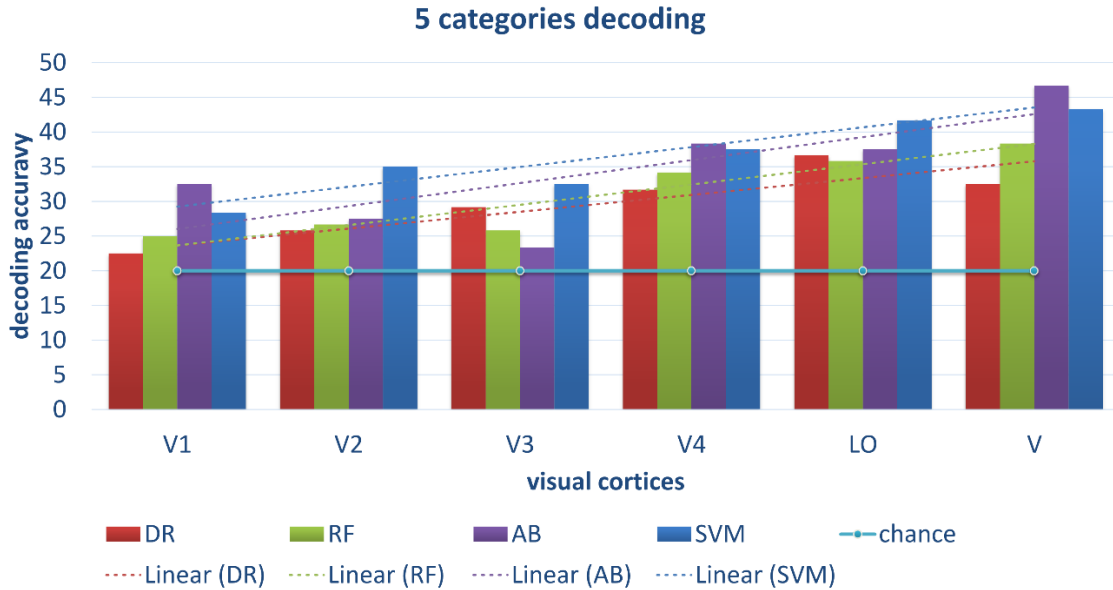

Figure 8. Decoding of 5 categories using conventional classifiers. Results demonstrated the phenomenon: distributed, hierarchical, and complementary representations about the semantic category in human visual cortices.

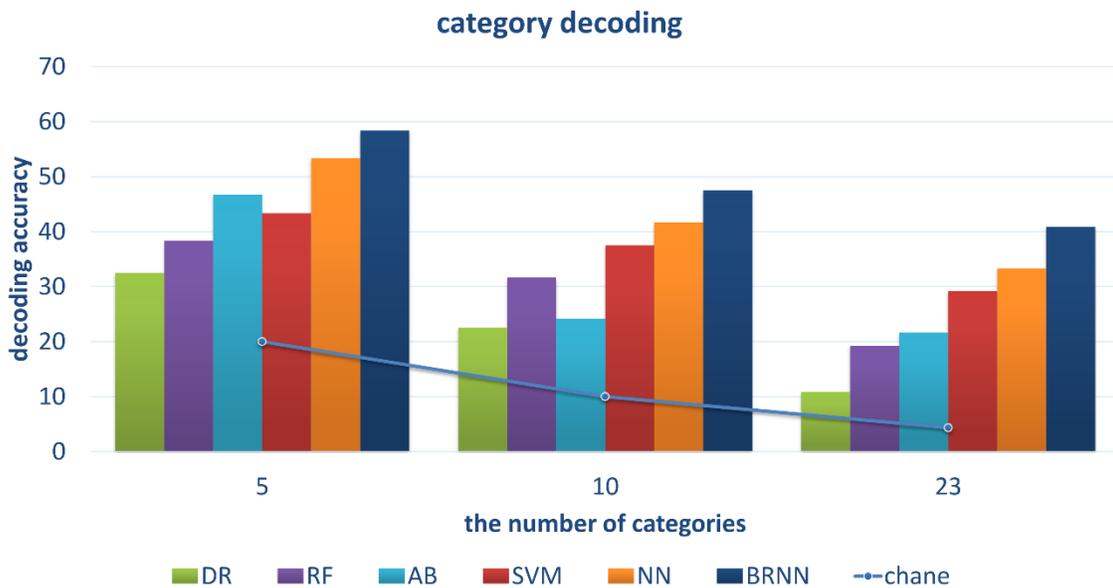

Figure 9. Qualitative decoding performance with comparison of different methods. Conventional methods and NN method can employ all visual cortices. However, the neural network with powerful nonlinear ability performed better. BRNN-based method with powerful nonlinear ability can also employ additional information (bidirectional information flows), leading to the best performance.

Table 1. Structure of the ResNet-50 model. All modules, layer names, corresponding channels, and feature size are presented. The channels become larger and feature sizes become smaller because of the down sampling.

| index | 1 | 2 | 3 | 4 | 5 | 6 | 7 | 8 | 9 | 10 | 11 | 12 | 13 | 14 | 15 | 16 | 17 | 18 |
|---|---|---|---|---|---|---|---|---|---|---|---|---|---|---|---|---|---|---|
| module | - | 1 | | | | 2 | | | | 3 | | | | | | 4 | | - |
| Name | conv1 | block1 | block2 | block3 | block1 | block2 | block3 | block4 | block1 | block2 | block3 | block4 | block5 | block6 | block1 | block2 | block3 | avgpool |
| channel | 64 | 64 | 256 | 256 | 256 | 512 | 512 | 512 | 512 | 1024 | 1024 | 1024 | 1024 | 1024 | 1024 | 2048 | 2048 | 2048 |
| feature size | 112×112 | 56×56 | 56×56 | 56×56 | 56×56 | 28×28 | 28×28 | 28×28 | 28×28 | 14×14 | 14×14 | 14×14 | 14×14 | 14×14 | 14×14 | 7×7 | 7×7 | 1×1 |

Table 2. Quantitative decoding performance with comparison of different methods for subject S1. BRNN-based method obtained about 11% and 5% improvement than the conventional traditional classifiers and NN method, respectively, which validated our proposed method and indicated the significance of bidirectional information flows.

| Category level | DR | RF | AB | SVM | NN | BRNN |
|---|---|---|---|---|---|---|
| 5 | 32.50 | 38.33 | 46.67 | 43.33 | 53.33 | 58.33 |
| 10 | 22.50 | 31.67 | 24.17 | 37.50 | 41.67 | 48.33 |
| 23 | 10.83 | 19.17 | 21.67 | 29.17 | 33.33 | 40.83 |

Table 3. Quantitative decoding performance with comparison of different methods for subject S2.

| Category level | DR | RF | AB | SVM | NN | BRNN |
|---|---|---|---|---|---|---|
| 5 | 29.17 | 37.50 | 35.00 | 34.17 | 38.33 | 42.50 |
| 10 | 11.67 | 21.67 | 17.50 | 26.67 | 33.33 | 38.17 |
| 23 | 10.83 | 19.17 | 18.33 | 18.33 | 21.67 | 26.67 |